\documentclass[final,nofootinbib,aps,pre,twocolumn,showpacs,groupaddress,preprintnumbers,floatfix]{revtex4-1}

\usepackage{dynlearn}

\newcommand{\Cg}{\ensuremath{C_g}\xspace}
\newcommand{\Cl}{\ensuremath{C_{L}}\xspace}
\newcommand{\Lohr}{L{\"o}hr\xspace}

\colorlet{cmu_color}{black}
\colorlet{cg_color} {blue}
\colorlet{cl_color} {red}
\colorlet{ee_color} {green!66!black}

\begin{document}

\def\ourTitle{Prediction and Generation of Binary Markov Processes:\\
Can a Finite-State Fox Catch a Markov Mouse?
}

\def\ourAbstract{Understanding the generative mechanism of a natural system is a vital component
of the scientific method. Here, we investigate one of the fundamental steps
toward this goal by presenting the minimal generator of an arbitrary binary
Markov process. This is a class of processes whose predictive model is well
known. Surprisingly, the generative model requires three distinct topologies
for different regions of parameter space. We show that a previously proposed
generator for a particular set of binary Markov processes is, in fact, not
minimal. Our results shed the first quantitative light on the relative
(minimal) costs of prediction and generation. We find, for instance, that the
difference between prediction and generation is maximized when the process is
approximately independently, identically distributed.
}

\def\ourKeywords{  stochastic process, hidden Markov model, \texorpdfstring{\eM}{epsilon-machine}, causal states, mutual information.
}

\hypersetup{
  pdfauthor={James P. Crutchfield},
  pdftitle={\ourTitle},
  pdfsubject={\ourAbstract},
  pdfkeywords={\ourKeywords},
  pdfproducer={},
  pdfcreator={}
}

\author{Joshua Ruebeck}
\email{ruebeckj@carleton.edu}
\affiliation{Department of Physics and Astronomy, Carleton College, One North College Street, Northfield, MN 55057}

\author{Ryan G. James}
\email{rgjames@ucdavis.edu}
\affiliation{Complexity Sciences Center and Physics Department, University of California at Davis, One Shields Avenue, Davis, CA 95616}

\author{John R. Mahoney}
\email{jrmahoney@ucdavis.edu}
\affiliation{Complexity Sciences Center and Physics Department, University of California at Davis, One Shields Avenue, Davis, CA 95616}

\author{James P. Crutchfield}
\email{chaos@ucdavis.edu}
\affiliation{Complexity Sciences Center and Physics Department, University of California at Davis, One Shields Avenue, Davis, CA 95616}

\date{\today}
\bibliographystyle{unsrt}

\title{\ourTitle}

\begin{abstract}

\ourAbstract

\vspace{0.1in}

\end{abstract}


\preprint{\sfiwp{17-07-XXX}}
\preprint{\arxiv{1707.XXXX}}

\title{\ourTitle}
\date{\today}
\maketitle

\setstretch{1.1}

\section{Introduction}
\label{sec:introduction}

Imagine a mouse being chased by a fox. Survival suggests that the mouse should
\emph{generate} a path that is difficult for the fox to predict. We might
imagine that the mouse brain is designed or trained to maximize the fox's
difficulty and, similarly, that the fox somehow has optimized the task of
\emph{predicting} the mouse's path. Are these two tasks actually distinct?  If
so, do there exist escape paths that are easier to generate than predict?
Every animal has limited computational resources and we might reasonably
suppose that the mouse has fewer than the fox. Given that mice clearly continue
to survive, we can ask whether this disparity in resources exists in tension
with the disparity in task-complexity---path-generation versus path-prediction.

In lieu of mouse paths, we consider the space of discrete stationary stochastic
processes---objects consisting of temporal sequences that span the range from
perfectly ordered to completely random. We then frame resource questions
quantitatively via hidden Markov model (HMM) representations of these
processes. We focus on two particular HMM representations of any given process:
the minimal predictive HMM---its computational mechanics' \eM
\cite{Crut12a}---and its minimal generative HMM. We then find two primary
measures of memory resource: \Cmu---defined as the \eM's
state-entropy---quantifies the cost of prediction, while \Cg---the state
entropy of the generative machine---quantifies the cost of generation.
Introduced over two and a half decades ago, the \eM predictive representation
is well studied and can be constructed for arbitrary processes~\cite{Shal98a}.
The generative machine offers more challenges, as it involves a nonconvex
constrained minimization over high-dimensional spaces. While there are several
known bounds on \Cg and restrictions on the construction of generative
HMMs~\cite{Lohr12, Lohr09b, Lohr09c, Hell65}, they have received significantly
less attention than the predictive case and, as a consequence, are markedly
less well understood.

The following presents the first construction of the minimal generators for an
arbitrary stationary binary Markov process. This allows for the analytic
calculation of \Cg, as well as other properties of generative models. These
models elucidate the differences between the tasks of generation and
prediction. The techniques introduced here should also lead to minimal
generators for other process classes.

\section{Models}
\label{sec:models}

We represent stochastic processes using edge-emitting (Mealy) \emph{hidden
Markov models} (HMMs). Such a representation is specified by a set of states, a
set of output symbols, a set of labeled transition matrices, and a stationary
distribution over states. We consider stationary processes so, assuming the
state transition structure is mixing, the invariant state distribution is
unique and is therefore redundantly determined from the labeled transition
matrices.

\begin{figure}
\centering
\includegraphics{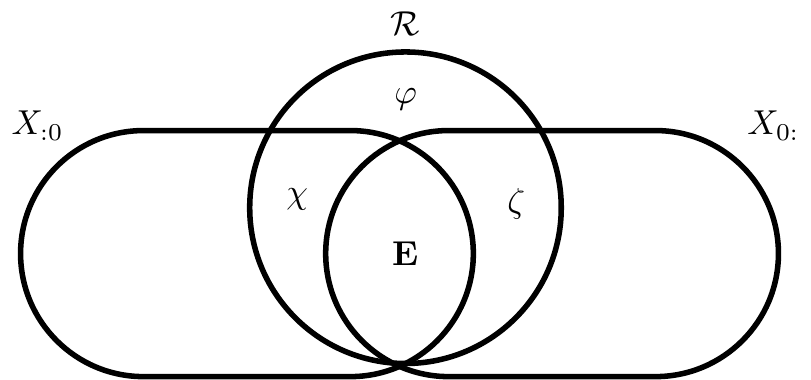}
 \caption{I-diagram~\cite{Yeun91a} of the $\Past - \AlternateState_0 - \Future$
	Markov chain \cite{Cove06a}. The state of a generating model shields the
	past and future, rendering them conditionally independent. This is
	reflected by the overlap \EE between the past and future being entirely
	captured (contained within) the system state entropy $\H{\AlternateState}$.
	The past and future further segment \AlternateState into the
	\emph{crypticity} \PC, \emph{gauge information} \GI, and \emph{oracular
	information} \OI; quantities whose interpretation is explored further in
	Ref.~\cite{Crut10a}.
  }
  \label{fig:correct_model}
\end{figure}

Clearly, not every HMM corresponds to any given process. If a model is to
correspond to a particular process, its states must yield \emph{conditional
independence} between the past and future. That is, the past $\MS{-\infty}{0}$
and future $\MS{0}{\infty}$ random variable chains yielded by a model must be
rendered independent by the model's current state $\AlternateState_0$;
information-theoretically, we have: $\I{\MS{-\infty}{0}: \MS{0}{\infty} |
\AlternateState_0} = 0$. The (unconditioned) mutual information $\EE =
\I{\MS{-\infty}{0} : \MS{0}{\infty}}$ between past and future is called the
\emph{excess entropy}. Among other uses, it is the amount of uncertainty about
the future one may reduce through knowledge of the past. Intuitively then, the
state of a correct model must ``capture'' \EE\ bits of information; see
Fig.\nobreakspace \ref {fig:correct_model}. (For brevity, the following suppresses infinite
variable indexes.)

There are an infinity of models for a given stochastic process. Depending on
context, certain models will have merits above those of others. The ability to
predict is one such context.

\section{Predictive Models}
\label{sec:prediction}

What is \emph{prediction}? Loosely speaking, prediction has to do with a
relation between two variables, one which we think of as input and the other as
output. In our context of stochastic processes, the input is the past \Past and
the output is the future \Future. By prediction we mean: Given some
instance of the past \past, the task is to yield the exact conditional
probability distribution $\Prob(\MS{0}{\ell} | \past)$ for any length $\ell$.

\subsection{\EM Construction}
\label{subsec:em_construction}

The minimal predictive model of a process \Process is known as its \eM and
its construction is straightforward. The theory of computational mechanics
provides a framework for the detailed characterization of \eMs in topological
and information-theoretic terms \cite{Crut12a}.

The kernel underlying this construction is the \emph{causal equivalence
relation} $\CausalEquivalence$. This is a relation over the set $\{\past\}$ of
semi-infinite pasts such that two pasts, $\past$ and $\pastprime$, belong to
the same equivalence class if their conditional futures agree:
\begin{align*}
  \past ~\CausalEquivalence~ \pastprime \iff \Prob(\Future | \past)
  = \Prob(\Future | \pastprime)
  ~.
\end{align*}
Each equivalence class is a state of the system, encapsulating in minimal form
the degree to which the past influences the future. Thus, we refer to the
classes as \emph{causal states} and denote by $\CausalState_t$ the causal state
at time $t$. The memory required by the \eM to \emph{implement} the act of
prediction is $\Cmu = \H{\CausalState}$---the \emph{statistical complexity}.
\footnote{This notion of memory applies in the ensemble setting. Single-shot or single-instance memory is also of interest and is studied in .}

Then, transitions between these states follow directly from the equivalence relation:
\begin{align*}
  T_{i,j}^k = \Prob(\MeasSymbol_0 = k, \CausalState_{1} = j | \CausalState_{0} = i)~.
\end{align*}

As previously stated, the excess entropy \EE is the amount of information
shared between past and future. The causal equivalence relation induces a
particular random variable \CausalState that ``captures'' \EE. Importantly, \EE
is not itself the entropy of a random variable. Thus, the causal-state random
variable cannot generally be of size \EE bits. We might then think of the
difference $\PC = \Cmu - \EE$, also known as the \emph{crypticity}, as the
\emph{predictive overhead} \cite{Maho11a}. It is an interesting fact that a
nonzero predictive overhead \PC is generic in the space of all processes.

\subsection{Binary Markov Processes}
\label{sec:bmc_em}

Let us now narrow our focus and construct the predictive models for the
particular class of binary Markov processes. More specifically, we consider all
stationary stochastic processes over the symbol set $\{0,1\}$ with the Markov property:
\begin{align*}
  \Pr(\MeasSymbol_0 | \MS{-\infty}{0}) = \Pr(\MeasSymbol_0 | \MeasSymbol_{-1})
  ~.
\end{align*}

\begin{figure}
\centering
\includegraphics{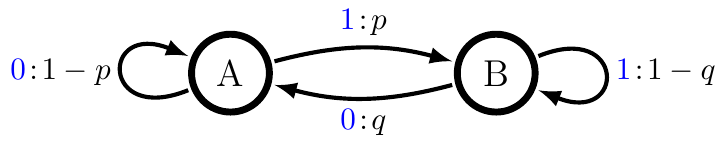}
 \caption{\EM for all binary Markov processes. Cases with $p = 1-q$ or $p=0$ or
	$q=0$ are single-state \eMs that are minimal in all respects: predictive or
	generative, entropic or dimensional.
	}
\label{fig:bmc_em}
\end{figure}

Applying the causal equivalence relation, we find that the causal state is
completely determined by the previous single symbol, a simple consequence of
the process' Markovity. This leads directly to the \eM in Fig.\nobreakspace \ref {fig:bmc_em}.

Its stationary state distribution is:
\begin{align*}
  \pi = \left[ \frac{1-q}{2-p-q}, \frac{1-p}{2-p-q} \right]
  ~.
\end{align*}

The informational properties of this class of processes---entropy rate, excess
entropy, and statistical complexity---can be stated in closed form:
\begin{align*}
  \hmu &= \pi_A \operatorname{H}\left(p\right)
  	+ \pi_B \operatorname{H}\left(q\right), \\
  \EE  &= \H{\pi} - \hmu, \\
  \Cmu &= \EE + \hmu
  ~,
\end{align*}
where $\operatorname{H}\left(p\right) = -(p \log{p}) + -((1-p) \log{(1-p)})$
denotes Shannon's binary entropy function \cite{Cove06a}. The simple relation
among these measures follows from the fact that any (nontrivial) binary Markov
process is also equivalent to a spin chain---a restricted class of Markov
chains \cite{Maho11a}.

\begin{figure}
  \centering
  \includegraphics{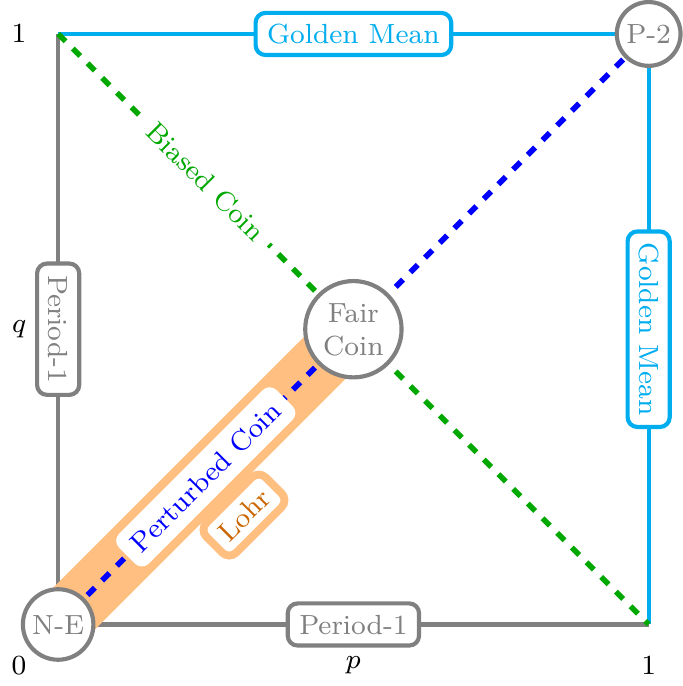}
   \caption{
    Process space spanned by binary Markov processes. When either $p = 0$ or
    $q = 0$, the process is constant, repeating $0$s or $1$s, respectively. In
    the limit $p = q = 0$ the process is nonergodic, realizing only one
	or the other of the two
    constant processes. When either $p = 1$ or $q = 1$, the expressed processes
    are known as \emph{Golden Mean Processes}, characterized by isolated $0$s
    or $1$s, respectively. Along the line $p = 1-q$ the process is a biased
    coin. Along the line $p = q$, the process is known as a perturbed coin,
    where states $A$ and $B$ each represent an oppositely biased coin and the
    process switches between the two biases based on the symbols just emitted.
  }
  \label{fig:process_space}
\end{figure}

This class of binary Markov processes spans a variety of structured processes,
summarized in Fig.\nobreakspace \ref {fig:process_space}. At the extremes of either $p = 0$ or $q
= 0$, we have a period-$1$ (constant) process. If either $p = 1$ or $q = 1$, we
have Golden Mean Processes, where $0$s or $1$s, respectively, occur in
isolation. If $p = 1 - q$, the process loses its dependence on the prior symbol
and it becomes a biased coin.

\section{Generative Models}
\label{sec:generative_models}

Let's now return to our original topic and describe the second type of process
representation---generative models. The only requirement of a generative model
is that it be able to correctly \emph{sample} from the distribution
$\Pr(\Future)$ over futures. More specifically, we require that, given any
instance \past of the past, the generative model yields a next symbol
$\MeasSymbol_0$ with the same probability distribution $\Pr(\MeasSymbol_0 |
\Past=\past)$ as specified by the process.

Note that, on the one hand, it may seem obvious that prediction subsumes generation. On the other, it is not so obvious how these two tasks might
prefer different mechanisms.

Like the \eM causal state, a generative state \AlternateState must also render
past and future conditionally independent. Importantly, as a consequence of the
causal equivalence relation \eMs are \emph{unifilar} which, when paired with
their minimality, implies that the causal states are functions of the prior
observables. Generative models, however, need not have this restriction.
Consequently, a given sequence of past symbols (finite or semi-infinite) may
induce more than one generative state.

Generative models are much less well understood than their predictive cousins.
This is due in large part to the lack of constructive methods for working with
and otherwise constructing them. This is why our results here, though
addressing only on a relatively simple class of processes, mark a substantial
step forward.

\begin{figure}
\centering
\includegraphics{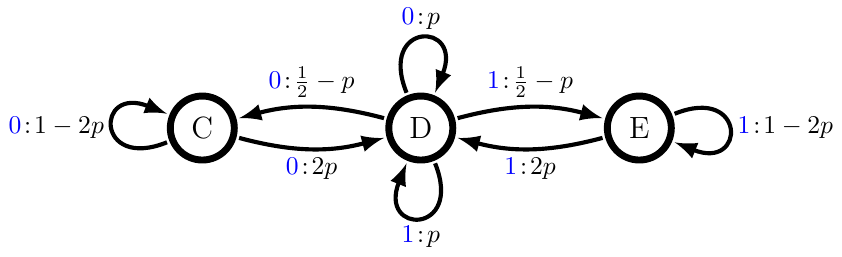}
 \caption{\Lohr model: A three-state HMM that generates the same process as
	that in Fig.\nobreakspace \ref {fig:bmc_em} when $0 \leq p = q \leq \nicefrac{1}{2}$. Its
	principle interest arises since it has a smaller state entropy than the
	\eM for a range of $p$ values: $\H{\AlternateState} \leq \Cmu$.
  }
\label{fig:lohr_machine}
\end{figure}

\section{\Lohr example}
\label{sec:lohr_example}

Let us now focus on a subclass of binary Markov processes, those for which $0 <
p = q < \nicefrac{1}{2}$; refer to the orange line in
Fig.\nobreakspace \ref {fig:process_space}.  Reference~\cite{Lohr09b} offers up a three-state HMM
generator for this class, which we refer to as the \emph{\Lohr model}; see
Fig.\nobreakspace \ref {fig:lohr_machine}. We see from the HMM that when probability $p$ is near
$\nicefrac{1}{2}$, the process is nearly \emph{independent, identically
distributed} (IID). An IID process has only a single causal state and therefore
zero statistical complexity, $\Cmu = 0$. However, for any deviation from $p =
\nicefrac{1}{2}$, the statistical complexity is a full bit, $\Cmu = 1$. Why is
it that a generator of a nearly IID process---that is, a nearly memoryless
process---still needs a full bit of memory?

The motivation for constructing this three-state model is that it might
concentrate the IID behavior into a single state and use the other states only
for those infrequent deviations that ``make up the difference''. And so, the
state-entropy may be reduced even though there are three states instead of two.
A priori it is not obvious that it is possible to yield the correct process in
this construction. It is, however, straightforward to check that the \Lohr
model produces the correct conditional statistics. It is a generator of the
processes. Note that in general it is sufficient to check these probabilities
for all words of length $2N-1$ where $N=\max(|\CausalState|,
|\AlternateState|)$ \cite[Corollary 4.3.9]{Uppe97a}.

We find that the \Lohr model has the stationary state distribution:
\begin{align*}
  \pi = \left[ \nicefrac{1}{2} - p, 2p, \nicefrac{1}{2} - p \right]
  ~.
\end{align*}
As noted, the statistical complexity $\Cmu = 1$ for $p$'s entire range. The
state entropy $\H{\AlternateState} = \H{\pi}$ of the \Lohr model is smaller
than $\Cmu=1$ for the range of values $p \in (0.38645\ldots, 0.5)$.
Importantly, this is sufficient to show that prediction and generation are
generally different tasks---they have different optimal solutions. This was
previously shown in Ref.~\cite{Lohr09b}. However, the question remained whether
or not the \Lohr model is \emph{minimal}. Surprisingly, though subsequent works
on generative complexity have appeared, to the best of our knowledge this
example is the only HMM published that is entropically smaller than the
(finite-state) \eM.

We will now construct the provably \emph{minimal} generator for these processes.
Further, we extend our analysis, not only to the range $p > \nicefrac{1}{2}$,
but to the entire $(p,q)$ domain of Fig.\nobreakspace \ref {fig:process_space}.

\begin{figure}
\centering
\includegraphics{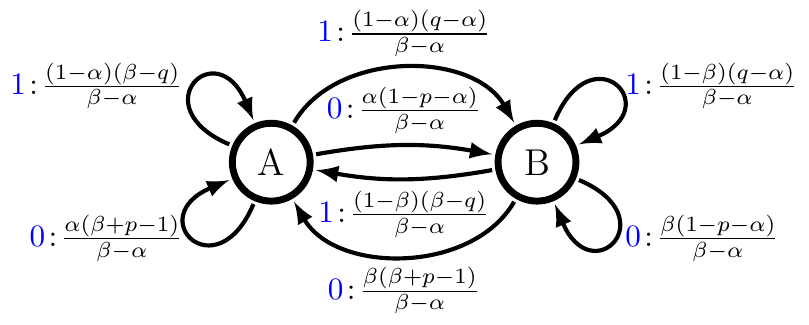}
 \caption{Parametrized HMM for the complete set of $2$-state machines that
	generate the space of binary Markov chain processes when
	$\alpha \leq \min\left\{ q, 1-p \right\}$ and
	$\beta \geq \max\left\{ q, 1-p \right\}$ and we assume $\alpha < \beta$.
	A second isomorphic class follows from the assumption $\beta < \alpha$.
  }
\label{fig:bmc_pqab}
\end{figure}

\subsection{Bounds}
\label{sec:bounds}

Recall that, for some $p$, the \Lohr model is entropically smaller than the \eM
and it achieves this while having three states instead of two. The important
point is that minimization of entropy in the generative context does not limit
the number of states in the same way as in the predictive one. (Recall that
among predictive models, the \eM is minimal in both entropy and state
number~\cite{Shal98a}.)

\begin{figure*}
\centering
\includegraphics{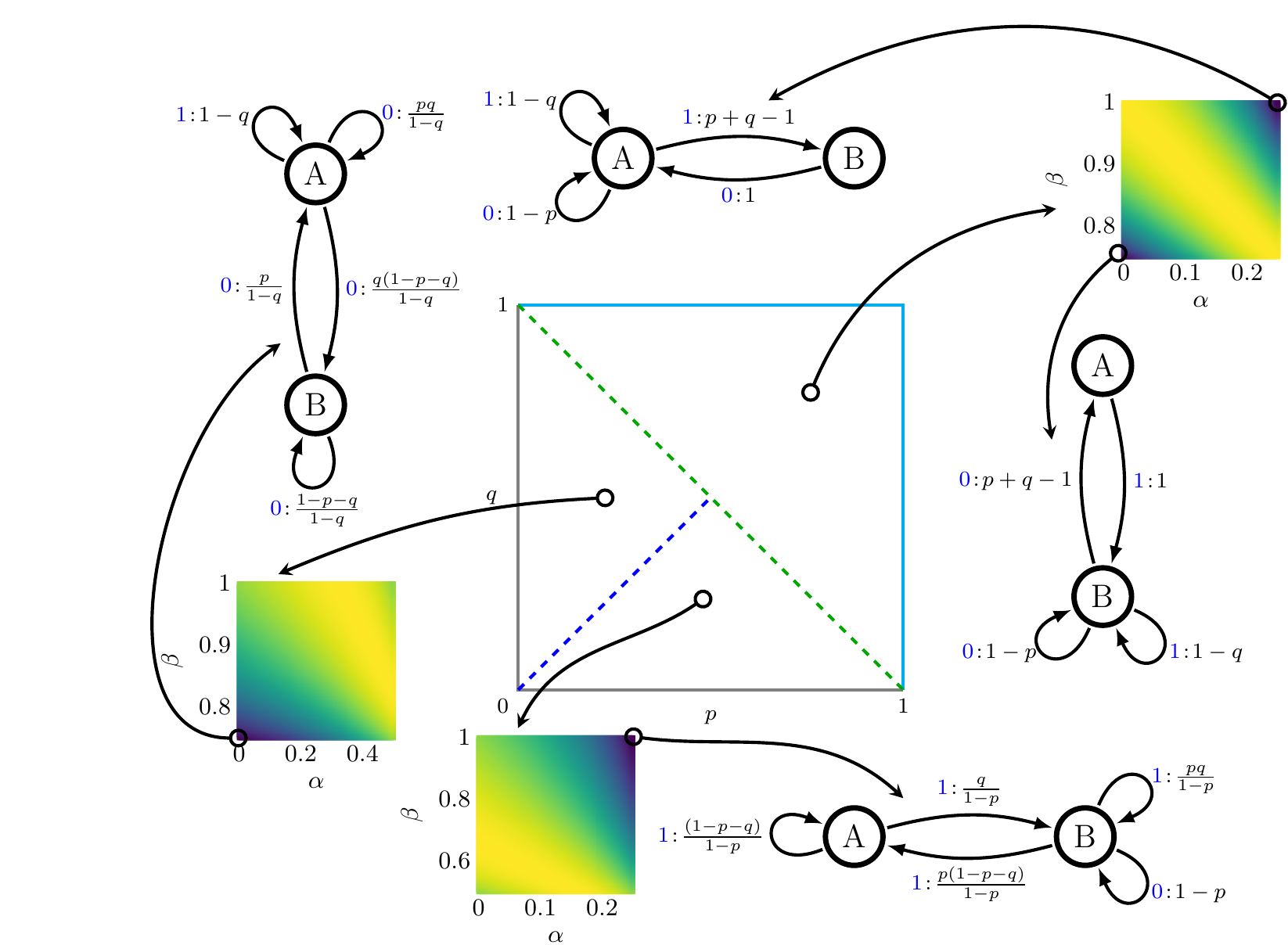}
 \caption{Two-parameter process space of binary Markov processes: Consider
	three points within this space. For each, there is a two-parameter model
	space. Within each model space we examine the model's state entropy and
	identify the global minima. We exhibit the corresponding HMMs. Topological
	changes in these minimal HMMs induce a three-region partition on process
	space.
  }
\label{fig:big_diagram}
\end{figure*}

A recent result shows that the maximum number of states in an entropically
minimal \emph{channel} $Z$ is $|Z| \leq \min \left\{ |X||Y|,
2^{\min\left\{|X|,|Y|\right\}} - 1 \right\}$, where $X$ and $Y$ are the channel
input and output processes~\cite{kumar2014exact}. Since a generative model is a
form of 	communication channel from the past to the future, we find that the
number of states of the minimal generative model is bounded by
$|\AlternateState| \leq \min \left\{ |\Past||\Future|,
2^{\min\left\{|\Past|,|\Future|\right\}} - 1 \right\}$. Of course, this result
is useless on its own: $|\Past|$ and $|\Future|$ are generically infinite.

This bound can be made practical by combining the data processing inequality
for exact common information $\G{X:Y}$~\cite{kumar2014exact} with the existence of the following two Markov chains~\cite{james2017trimming}:
\begin{align*}
  &\Past - \FCS - \RCS - \Future \ , \, \textrm{and} \\
  &\FCS - \Past - \Future - \RCS
  ~.
\end{align*}
We denote forward- and reverse-time causal states $\FCS$ and $\RCS$, respectively. Combined, these tell us that $\G{\Past:\Future} = \G{\FCS:\RCS}$. Therefore, the bound can be tightened to $|\AlternateState| \leq \min \left\{ |\FCS||\RCS|, 2^{\min\left\{|\FCS|,|\RCS|\right\}} - 1 \right\}$. This is a
particularly helpful application of causal states.

\subsection{Binary Markov Chains}
\label{sec:bmc_gen}

In the particular case of processes represented by binary Markov chains, the
reverse process is also represented by a binary Markov chain. And so, both
$|\FCS| = 2$ and $|\RCS| = 2$. From the above bounds, we find that
$|\AlternateState| \leq 3$. Closely following the proof in
Ref.~\cite{kumar2014exact}, one can then show that no three-state
representation is minimal. And, since a single state model can only represent
IID processes, this leaves only models with two-states as the possible minimal representations.

We begin with the assumption that an observation $\MeasSymbol_0$ maps
stochastically to a state $\AlternateState_0$, which then stochastically maps
to a symbol $\MeasSymbol_1$. Constraining this pair of channels to produce
observations $\MeasSymbol_0$ and $\MeasSymbol_1$ consistent with the binary
Markov chain yields the parametrized hidden Markov model found in
Fig.\nobreakspace \ref {fig:bmc_pqab}. (Appendix \ref{sec:derivation} gives the background calculations.)

For each point $(p,q)$ in the binary Markov \emph{process}-space, we now have a
two-parameter \emph{model}-space of HMMs, specified by $(\alpha, \beta)$. The
constraint that conditional probabilities be between zero and one restricts our
model-space parameters to a rectangle $\alpha \leq \min\left\{q, 1-p \right\}$
and $\beta \geq \max\left\{ q, 1-p \right\}$. One can now compute the state
entropy within this constrained model-space and identify the minima.

Since the entropy is concave in $\alpha$ and $\beta$ and the allowable regions
in $(\alpha, \beta)$-space are convex (rectangles), it is sufficient to search
for local minima along the boundary.

Figure\nobreakspace \ref {fig:big_diagram} illustrates this for three different points in process
space. We find that at each of the points $(p=\nicefrac{1}{4},
q=\nicefrac{1}{2})$ and $(p=\nicefrac{1}{2}, q=\nicefrac{1}{4})$, there is a
single global minimum. For the point $(p=\nicefrac{3}{4}, q=\nicefrac{3}{4})$,
we find that there are two minima equivalent in value, but corresponding to
nonisomorphic HMMs. Both representations are biased toward producing a periodic
sequence, with fluctuations interjected at different phases of the period.

In this way, one can discover the minimal generator for any binary Markov
chain. Examining these minimal topologies at each point, we find that
process-space is divided into three triangular regions with
topologically-distinct generators. This is a somewhat surprising contrast with
the fact that this model class requires only one predictive topology.

\begin{figure}
\centering
\includegraphics{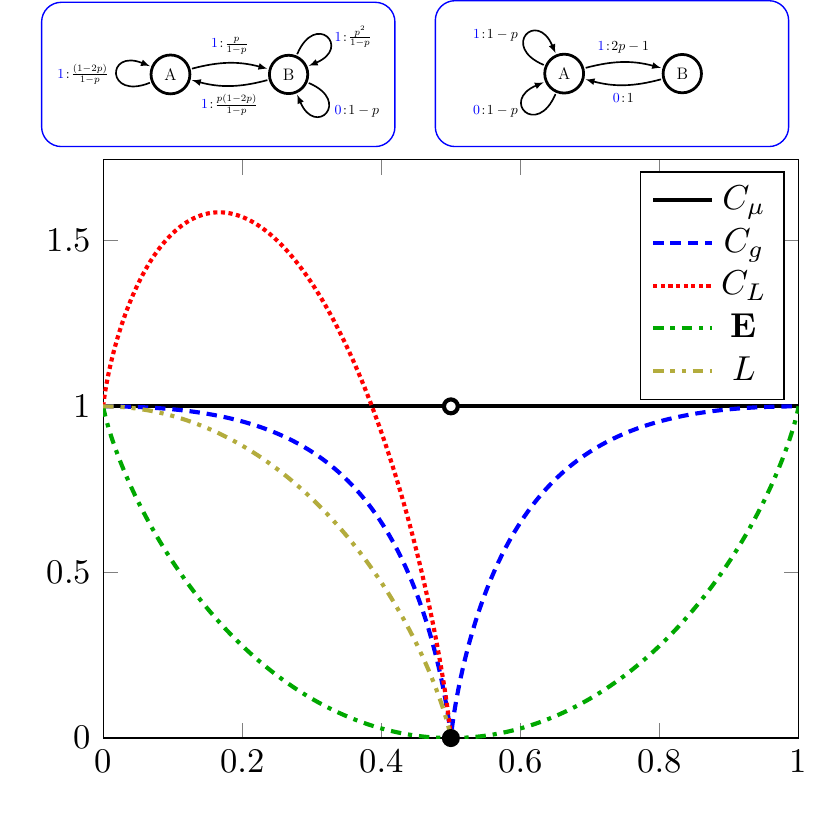}
 \caption{State entropy of various models of the perturbed coin: The excess
	entropy \EE is the amount of information any model of a process must
	possess. A stronger lower bound $L$ claimed by \Lohr is also plotted.
    Entropies of the three models: \Cmu for the \eM, \Cl for \Lohr's model, and
	\Cg for the generative model. While \Cl is less than \Cmu for some values
	of $p$, \Cg is less than both \Cmu and \Cl everywhere.
  }
\label{fig:perturbed_coin}
\end{figure}

Let us briefly return to the restricted process previously considered---the
perturbed coin. We may now quantitatively compare the three state-entropies of
interest. In Fig.\nobreakspace \ref {fig:perturbed_coin} we see that the statistical complexity
$\Cmu = 1$ everywhere, but at $p = \nicefrac{1}{2}$, where it vanishes, $\Cmu =
0$. The \Lohr model's state-entropy $C_L$ falls below \Cmu, but only for a
subset of $p$ values. However, the generative complexity \Cg (a smooth
function) is everywhere less than both \Cmu \emph{and} $C_L$. (The generative
models for $p < \nicefrac{1}{2}$ and $p > \nicefrac{1}{2}$ are shown above.)
This shows that the proposed \Lohr model is not the generative model for any
value of $p$.

As implied by the conditional independence requirement, the excess entropy \EE
remains a lower bound on each of these state-entropies. \Lohr~\cite{Lohr09b}
constructed a tighter lower bound, denoted $L$, on any model of the perturbed
coin. We see that \Cg is slightly larger than this bound. It may be useful to
generalize this lower bound for other processes.

\begin{figure}
\centering
\includegraphics{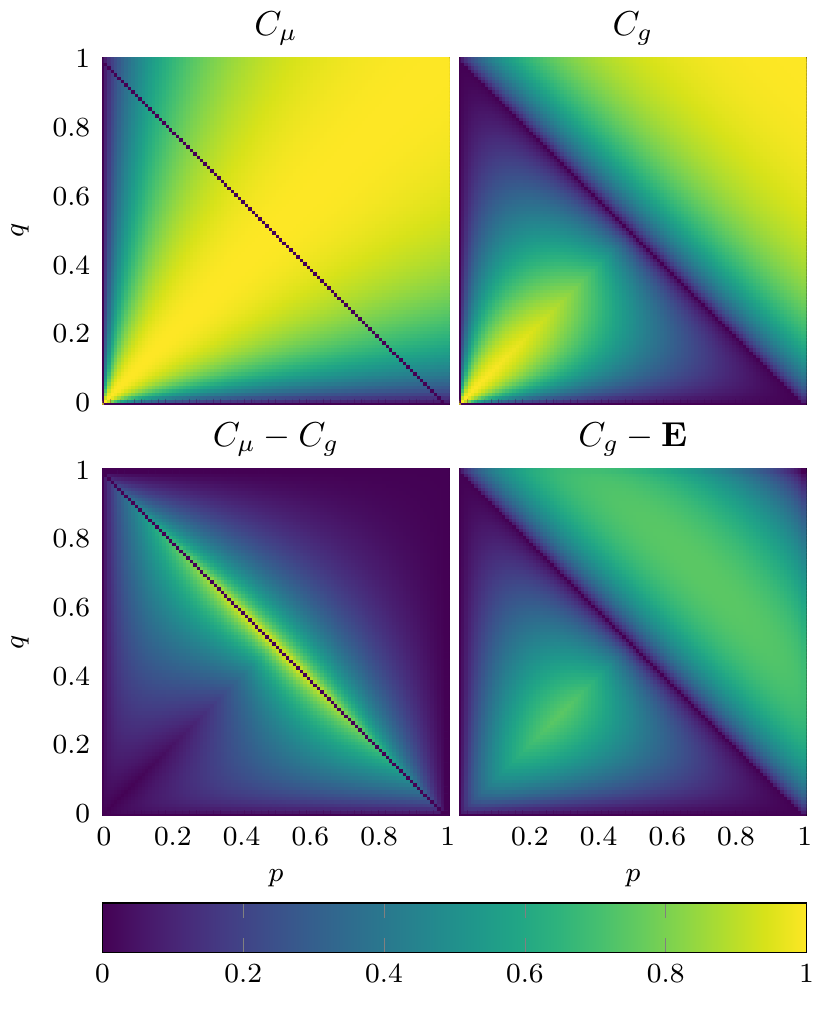}
 \caption{State complexity of the two canonical models: \eM and generative
	machine. The predictive overhead, $\Cmu - \Cg$, quantifies the information
	required to enable prediction above and beyond generation. The generative
	overhead, $\Cg - \EE$, quantifies the amount of information a model of a
	process requires beyond that minimally required by the observable
	correlations.
  }
\label{fig:cmu_cg}
\end{figure}

The minimal generators are defined over all of $(p,q)$-space. We can compare
the cost \Cmu of prediction with the cost \Cg of generation and the information
necessarily captured by a model---the excess entropy \EE. This comparison is
seen in Fig.\nobreakspace \ref {fig:cmu_cg}.

Focusing on the upper two panels of Fig.\nobreakspace \ref {fig:cmu_cg}, we see that both \Cmu
and \Cg display $p \leftrightarrow q$ symmetry. Furthermore, \Cg has a
discontinuous derivative along this line of symmetry, but only in the southwest
(SW).

For \Cmu, the line $p+q=1$ is special in that this lines marks a causal-state
collapse---two causal states merge into one under the equivalence relation.
For \Cg, however, this line marks a qualitative change in behavior (SW versus
NE). Since the generative complexity is lower semi-continuous~\cite{Lohr12}, we
know that a predictive gap $\Cmu - \Cg$ must exist around this line.

The lower two panels of Fig.\nobreakspace \ref {fig:cmu_cg} suggest that the costs of generation
and of prediction may have different causes. The parameters for which $\Cg -
\EE$ is high are disjoint from those where $\Cmu - \Cg$ is high.  \Cg is high
when $p$ and $q$ are correlated (near the $p$-$q$ symmetry line), but only for
$p ~ q < \nicefrac{1}{2}$. In the other half of parameter space, \Cg is high
when $p$ and $q$ are anti-correlated and away from the causal collapse. In
contrast, \Cmu is high exclusively near the line of causal collapse.

\section{Conclusion}
\label{sec:conclusion}

We presented the minimal generators of binary Markov stochastic processes.
Curiously, the literature appears to contain no other examples of generative
models, for processes with finite-state \eMs. And so, our contribution here is
a substantial step forward. It allows us to begin to understand the difference
between prediction and generation through direct calculation. It also opens
these new models to analysis by a host of previously developed techniques
including the information diagrams presented here.

To put the results in a larger setting, we note that HMMs have found
application in many diverse settings, ranging from speech recognition to
bioinformatics. And so, there are many reasons to care about the states and
information-theoretic properties of these models, some obvious and some not.
It is common to imbue a state with greater explanatory power than, say, a
random variable that merely exhibits the correct correlations for the
observables at hand. For instance, we may seek \emph{independent} means of
determining the state. Whether or not this is appropriate, the fact remains
that the different tasks of prediction and generation are associated with
different kinds of state, each with different kinds of explanatory usefulness.
This distinction seems to us to be rarely if ever made in HMM applications.

The concept of model state is central, for example, in model selection. A
simple and common method for selecting one model over another is through
application of a penalty related to the number of states (or entropy thereof)
\cite{Akai77a}. Since the predictive model will never have a lower entropy than
the corresponding generative one, an entropic penalty should never yield the
predictive model, however a state-number penalty might. Similarly, in model
parameter inference, if one distinguishes between the predictive and generative
classes, the maximum likelihood estimated parameters will differ between the
two classes.

Finally, we close by drawing out the consequences for fundamental physics.
Understanding states bears directly on thermodynamics. Landauer's Principle
states that erasing memory comes at a minimum, unavoidable cost---a heat
dissipation proportional to the size of the memory erased \cite{Land61a}. One
can consider HMMs as abstract representations of processes with memory (the
state) that must be modified or erased as time progresses. Applying Landauer's
Principle assigns thermodynamic consequences to the HMM time evolution. Which
HMM (and corresponding states) is appropriate, though? We now see that
prediction and generation, two very natural tasks for a thermodynamic system to
perform, actually deliver two different answers. It is important to understand
how physical circumstances relate to this choice of task---it will be expressed
in terms of heat.

\section*{Acknowledgments}
\label{sec:acknowledgments}

We thank the Santa Fe Institute for its hospitality during visits, where JPC is
an External Faculty member. This material is based upon work supported by, or
in part by, John Templeton Foundation grant 52095, Foundational Questions
Institute grant FQXi-RFP-1609, and the U. S. Army Research Laboratory and the
U. S. Army Research Office under contracts W911NF-13-1-0390 and
W911NF-13-1-0340. JR was funded by the 2016 NSF Research Experience for
Undergraduates program.

\clearpage
\begin{center}
\large{Supplementary Materials}\\
\emph{Prediction and Generation of Binary Markov Processes:\\
Can a Finite-State Fox Catch a Markov Mouse?}\\
Joshua Ruebeck, Ryan G. James, John R. Mahoney, and James P. Crutchfield
\end{center}

\setcounter{equation}{0}
\setcounter{page}{1}
\setcounter{section}{0}
\makeatletter
\renewcommand{\theequation}{S\arabic{equation}}

These supplementary materials give additional information on the processes and
models analyzed in the main body. They cover the relationship between causal
states and L{\"o}hr states, process information diagram analysis, including
crypticity and oracular information, and the derivation of the general,
parametrized binary Markov process model itself.

\section{Causal States and L{\"o}hr States}

Since the process considered in the main article is Markov order $R = 1$ and
its causal states are in one-to-one correspondence with the symbol last seen,
we can compactly represent the relation between pasts, causal states, and L{\"o}hr
states. The mapping from causal states $A$ and $B$ (or observation symbols $0$
and $1$) to states $C$, $D$, and $E$ of the L{\"o}hr presentation is given by:
\begin{align}
\Pr(\AlternateState | \CausalState)
  = \bordermatrix{
    ~   & C    & D  & E    & \cr
    A/0 & 1-2p & 2p & 0    & \cr
    B/1 & 0    & 2p & 1-2p
  }
\label{eq:lohr_map}
\end{align}

For example, if the last symbol was $x = 0$, this induces causal state $A$.
The corresponding L{\"o}hr states occur with respective probabilities $\Pr(C,D,E) =
(1-2p, 2p, 0)$.

\section{Information Diagram Analysis}
\label{sec:idiagram}

\begin{figure}
\centering
\clearpage{}\includegraphics{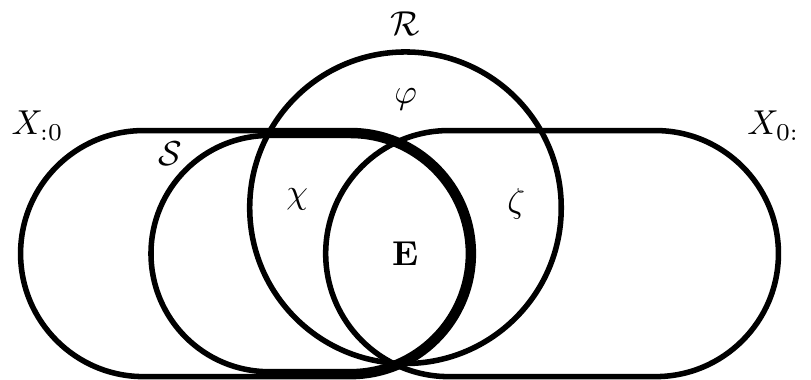}
\clearpage{}
\caption{I-diagram depicting how different kinds of information are shared
	among a process' past $\Past$, future $\Future$, (forward) causal states
	$\CausalState$, and generative states $\AlternateState$. Several
	information atoms are labeled: the process' past-future mutual information
	or excess entropy $\EE$, the crypticity $\PC$, the oracular information
	$\OI$, and the gauge information $\GI$.
	}
\label{fig:idiagram}
\end{figure}

The information diagram~\cite{Yeun91a} is a tool that has become increasingly
useful for analyzing and characterizing the information content of process
presentations~\cite{james2017trimming}. It gives a visual representation of
multivariate information-theoretic dependencies. It arises from a duality
between information measures and set theory: set-intersection corresponds to
mutual information and set-union corresponds to joining distributions.  Here, we
analyze several information measures introduced in Ref.~\cite{Crut10a} for the
processes considered in the main article; the associated I-diagram is depicted
in Fig. \ref{fig:idiagram}.

A generic model state $\AlternateState$ has nonzero intersection (mutual
information) with both the process' past $\Past$ and future $\Future$:
$\I{\AlternateState:\Past} > 0$ and $\I{\AlternateState:\Future} > 0$.
Importantly, the state information of a model that correctly generates a
process---a process presentation---must intersect the past-future union, which
completely contains the past-future intersection.  We might say that the state
of the model ``captures'' this atom---the excess entropy $\EE =
\I{\Past:\Future}$. Any model that does not capture at least $\EE$ of the past
and future cannot correctly generate the process. (Recall that if a model is
predictive, then it is also generative.)

The \eM is simple not only in the sense that it is constructible, but it is
informationally simple as it appears in the information diagram as well.  This
simplicity follows from the fact that causal states are functions of the past.
As a result, \eM state information has no intersection with the future beyond
that of that past's---it contains no, what we call, \emph{oracular} information
$\OI = \I{\AlternateState:\Future|\Past}$. It also contains no information
outside the past-future union---it contains not \emph{gauge} information $\GI =
\H{\AlternateState|\Past,\Future}$. Beyond the excess entropy atom, it has only
one potential region---what we call the process' \emph{crypticity} $\PC_{\eM} =
\Cmu - \EE$, where $\Cmu = \H{\CausalState}$ is the process' \emph{statistical
complexity}.

The situation is richer for general representations, including generative ones;
denote them $g$. The more general definition of crypticity is $\PC_g =
\I{\AlternateState : \Past | \Future}$. A general representation may have
oracular and gauge information. And, it may have a crypticity greater than,
equal to, or less than the \eM crypticity.

Generative models are restricted in two ways.  First, their crypticity is never greater than the \eM's crypticity: $\PC_g \leq \PC_{eM}$. This follows straightforwardly since if it were greater, then the \eM would necessarily have smaller state entropy, leading to a contradiction. Further, the sum of the generative model's crypticity, gauge information, and oracular information must be less than the \eM's crypticity:
\begin{align}
  \PC_g + \GI_g + \OI_g \leq \PC_{eM}
  \label{eq:gen_vs_em}
\end{align}

\begin{figure}
\centering
\includegraphics{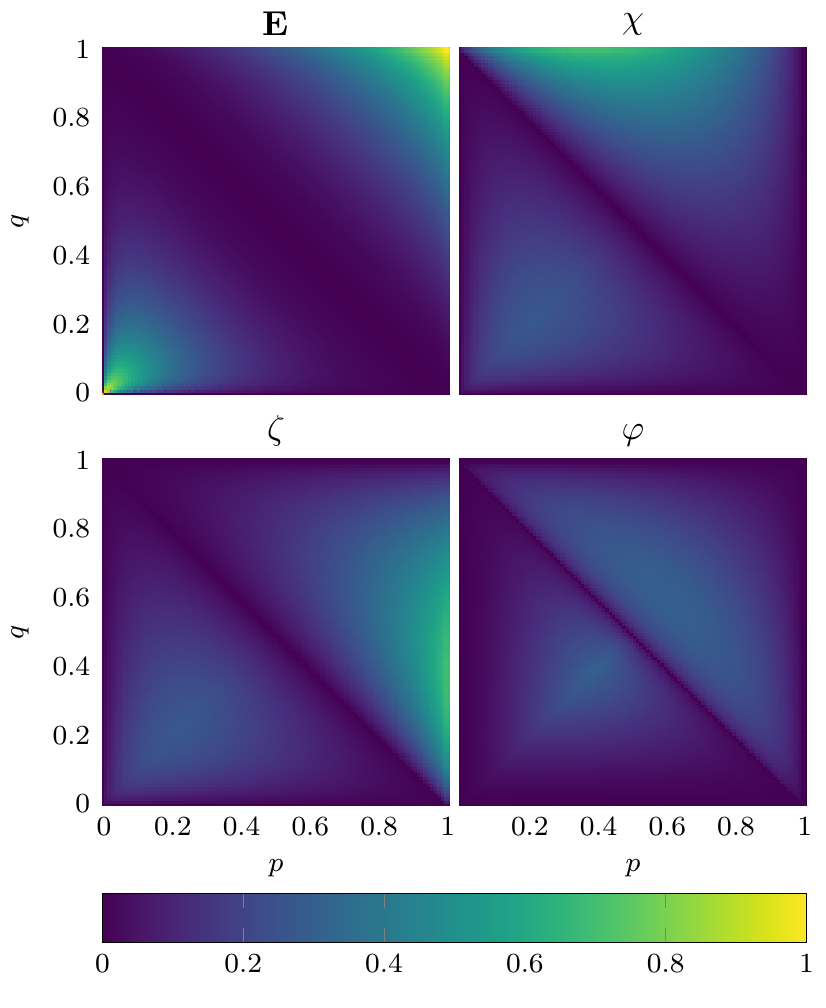}
 \caption{Decomposition of the generative state information \Cg into excess
	entropy $\EE$, crypticity $\PC$, oracular information $\OI$, and gauge
	information $\GI$ over process space $(p,q)$.
  }
\label{fig:cg_synccontrol}
\end{figure}

\begin{figure}
\centering
\includegraphics{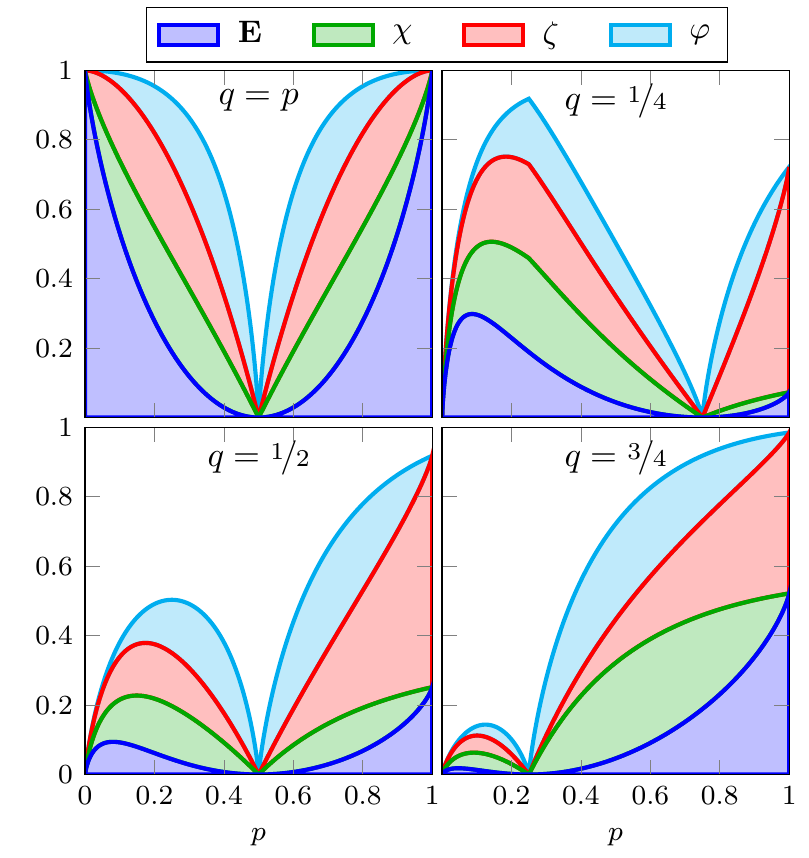}
 \caption{Decomposition of \Cg for several slices of process space, revealing
	more directly their functional dependencies.
 }
 \label{fig:cg_slices}
\end{figure}

Second, appealing to our newly introduced generative models for binary Markov
chains, we can compare the generative-model crypticity to that of the \eM. A
generative model distinct from the \eM must have nonzero oracular
information~\cite{kamath2010new}. Also, a generative model may only have
nonzero gauge information if it has both crypticity and oracular
information~\cite{kamath2010new}:
\begin{align}
  \GI_g > 0  \implies
    \begin{cases}
      \PC_g > 0\\
      \OI_g > 0
    \end{cases}
  ~,
\end{align}
while still satisfying Eq.\nobreakspace \textup {(\ref {eq:gen_vs_em})}. Effectively, this means that gauge
information can be ``minimized'' away, unless it supports the existence of both
cryptic and oracular information.

For our parametrized generative models of binary Markov chains we find that \Cg
decomposes as shown in the two mosaics in Figs.\nobreakspace \ref {fig:cg_synccontrol} and\nobreakspace  \ref {fig:cg_slices}. The gauge information $\GI_g$ is,
generally, a rather small portion of \Cg, though it is largest in the same
regions as $\Cg - \EE$. Both the crypticity $\PC_g$ and the oracular
information $\OI_g$ are large along the $q=1$ and $p=1$ edges, respectively.
This implies that $\AlternateState$ ``leans left'' when $q \approx 1$ and
``leans right'' when $p \approx 1$. This relationship flips if we use the
alternative, equivalent model in the $p > 1 - q$ half of the process space.

Note that while gauge information is not required of generative models, it is
present in all of the binary Markov chain generators, except along boundary and
causal collapse lines. Surveys of other processes suggest that the presence of
gauge information may be somewhat rare.

\section{Parametrized Binary Markov Process HMM: Derivation}
\label{sec:derivation}

We derive the most generic parametrized, two-state HMM of binary Markov
processes. The target binary Markov process is represented by the conditional
probability matrix:
\begin{align}
  \Pr(X_t | X_{t-1}) = \bordermatrix {
    ~ & 0     & 1     \cr
    0 & 1 - p & p     \cr
    1 & q     & 1 - q
  }
\label{eq:binary_chain_trans_matr}
\end{align}

As an intermediate step to deriving the full HMM, we find
the conditional probabilities involving the machine states relative
to the symbols emitted on the transitions to and from a state. We
write down two template matrices, denoting the states $A$ and $B$
and the random variable over them $\AlternateState$:
\begin{align}
  \Pr(\MeasSymbol_{t} | \AlternateState)= \bordermatrix{
    ~ & 0      & 1          \cr
    A & \alpha & 1 - \alpha \cr
    B & \beta  & 1 - \beta
  }
\label{eq:param_model}
\end{align}
and:
\begin{align}
  \Pr(\AlternateState | \MeasSymbol_{t-1}) = \bordermatrix {
    ~ & A      & B          \cr
    0 & \gamma & 1 - \gamma \cr
    1 & \delta & 1 - \delta
  }
  ~.
\label{eq:param_model2}
\end{align}

there is ambiguity here in labeling the states. If we switch the rows of
Eq.~(\ref{eq:param_model}) with each other and the columns of
Eq.~(\ref{eq:param_model2}) with each other, then we obtain different matrices
that describe the same model. Only the the state labels have been swapped. To
avoid double-counting such machines, we restrict $\alpha \leq \beta$.

Multiplying out this factorization and requiring it to be consistent
with the target provides two constraints:
\begin{align}
  \gamma = \frac{\beta + p - 1}{\beta-\alpha} \quad, \quad
  \delta = \frac{\beta-q}{\beta-\alpha}.
  \label{eq:gamma-delta}
\end{align}
(We keep the denominators as $\beta-\alpha$ so that both the numerator and
denominator are positive.) We ask that all these be valid probabilities.
Requiring $0\leq \gamma,\delta\leq 1$ and substituting
Eq.~\eqref{eq:gamma-delta} yields the constraints:
\begin{align*}
\alpha & \leq \min\{q,1-p\},\\
  \beta & \geq \max\{q,1-p\}
  ~.
\end{align*}

\begin{figure}
\centering
\includegraphics{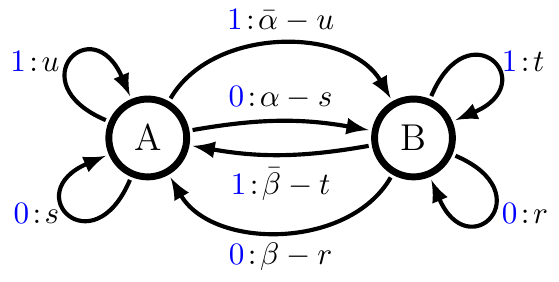}
 \caption{As an intermediate step, we construct a skeleton for the
    fully general HMM. $r,s,t,u$ are temporary variables to be solved
    for eventually. The other transitions are labeled using
    the fact that, for example, the probabilities of the paths coming
    from state $A$ and emitting a $0$ must sum to $\alpha$ in order to
    agree with Eq.~\ref{eq:param_model}.
	}
\label{fig:bmc_pqrstu}
\end{figure}

Now, consider the incomplete HMM that defines the helper variables $r,s,t,u$ in Fig.\nobreakspace \ref {fig:bmc_pqrstu}. To determine the latter's values, we evaluate the probabilities of a string (a) being generated by this incomplete HMM:
\begin{align}
\Pr(01) & = \Pr(01|A)\Pr(A) + \Pr(01|B)\Pr(B) \nonumber \\
        & = [s(1-\alpha) +(\alpha-s)(1-\beta)]\Pr(A) \nonumber \\
        & \qquad + [(\beta-r)(1-\alpha)+r(1-\beta)]\Pr(B)
		\label{eq:pr01a}
\end{align}
and (b) being generated by the Markov process
described in Eq.~(\ref{eq:binary_chain_trans_matr}):
\begin{align}
  \Pr(01) &= \Pr(1|0)\Pr(0) \nonumber \\
          &= p\Pr(0)
		  ~.
\label{eq:pr01b}
\end{align}
We used the string $01$ as our first case. And, $\Pr(0)$,
$\Pr(1)$, $\Pr(A)$, and $\Pr(B)$ are given by the stationary
distributions over the symbols and states, respectively:
\begin{align*}
  \Pr(0) &= \frac{q}{p+q}\\
  \Pr(1) &= \frac{p}{p+q}\\
  \Pr(A) &= \frac{q\gamma + p\delta}{p+q}\\
  \Pr(B) &= \frac{q(1-\gamma)+p(1-\delta)}{p+q}
  ~.
\end{align*}

Setting Eqs.~(\ref{eq:pr01a}) and Eq.~(\ref{eq:pr01b}) equal to one another
constrains $r,s,t,u$. To fully specify all four, though, we need
three more independent equations. We obtain these by evaluating the
probabilities of the strings $11$, $110$, and $000$ in a similar
fashion. Respectively, these yield:
\begin{align*}
(1-q)\Pr(1) &= [u(1-\alpha)+(1-\alpha-u)(1-\beta)]\Pr(A) \\
            & \quad + [t(1-\beta)+(1-\beta-t)(1-\alpha)]\Pr(B)
  ~, \\
q(1-q)\Pr(1) &= [u((1-\alpha-u)\beta+u\alpha)\nonumber\\
             & \quad +(1-\alpha-u)(t\beta+(1-\beta-t)\alpha)]\Pr(A)\nonumber\\
             & \quad + [t(t\beta+(1-\beta-t)\alpha)\nonumber\\
             & \quad +(1-\beta-t)((1-\alpha-u)\beta+u\alpha)]\Pr(B)
  ,~\text{and} \\
(1-p)^2\Pr(0) &= [s((\alpha-s)\beta+s\alpha)\nonumber\\
              & \quad +(\alpha-s)(r\beta+(\beta-r)\alpha)]\Pr(A)\nonumber\\
              & \quad +[r((\beta-r)\alpha+r\beta)\nonumber\\
              & \quad +(\beta-r)((\alpha-s)\beta+s\alpha)]\Pr(B) ~.
\end{align*}
Solving this system yields:
\begin{align*}
  r & =\frac{\beta(1-p-\alpha)}{\beta-\alpha} ~,\\
  s & =\frac{\alpha(\beta+p-1)}{\beta-\alpha} ~,\\
  t & =\frac{(1-\beta)(q-\alpha)}{\beta-\alpha} ~,~\text{and}\\
  u & =\frac{(1-\alpha)(\beta-q)}{\beta-\alpha}
  ~.
\end{align*}
Finally, we substitute these into Fig.~\ref{fig:bmc_pqrstu} to recover the
parametrized HMM the main article introduced in Fig. \ref{fig:bmc_pqab}.

\end{document}